\documentclass{aip-cp}
\usepackage{comment}
\usepackage[numbers]{natbib}
\usepackage{rotating}
\usepackage{graphicx,epsfig}

\begin{document}

\title{Ergodic statistical models: entropic dynamics and chaos}

\author[aff1]{Ignacio S. Gomez\corref{cor1}}
\author[aff1]{M. Portesi}
\eaddress{portesi@fisica.unlp.edu.ar}

\affil[aff1]{IFLP, UNLP, CONICET, Facultad de Ciencias Exactas, Calle 115 y 49, 1900 La Plata, Argentina}
\corresp[cor1]{nachosky@fisica.unlp.edu.ar}

\maketitle

\begin{abstract}
We present an extension of the ergodic,
mixing and Bernoulli levels of the ergodic hierarchy in dynamical systems, the information geometric ergodic hierarchy (IGEH), making use of statistical models on curved manifolds in the context of information geometry. We discuss the $2\times 2$ Gaussian Orthogonal Ensembles (GOE) within a
$2D$ correlated model.
For $r$ vanishingly small, we find that GOE belong to
the information geometric (IG) mixing level having a maximum negative value of scalar curvature.
Moreover, we propose a measure of distinguishability for the family of
distributions of the $2D$ correlated model that results to be an upper bound of the IG correlation.
\end{abstract}


\section{INTRODUCTION}
Historically, several fundamental disciplines like statistical mechanics and thermodynamics in physics, have turned out to be adequately treated as theories of inference where the conclusions are derived from available information about the system and making use of a method of reasoning that involves  probability theory. In this sense, one of the most important methods of inference is the \emph{maximum entropy method} (ME method) where a rule is given for obtaining the distribution $p$ that represents the best knowledge of the system and corresponds to maximum uncertainty, as measured by a functional $S[p]$, constrained by the available information \cite{jaynes}.
In particular, the so-called \emph{MaxEnt} method results when the entropy $S[p]$ is chosen to be the Shannon--Gibbs one.

Given a well-established rule of inference like MaxEnt and generally other ME methods \cite{giffin}, one can study the dynamics of the system within that framework.
The approach proposed by Caticha et al. \cite{caticha1,caticha2}, known as \emph{Entropic Dynamics} (ED), is to consider that the system moves irreversibly and continuously along the entropy gradient in a curved statistical manifold whose elements are probability distributions.
Curved statistical manifolds are the subject of study of the \emph{information geometry} (IG), and they have associated the Fisher--Rao metric \cite{rao} which in turn is linked to the concepts of entropy and Fisher information \cite{rao, amari}. Thus ED is a theoretical framework that arises
from the combination of ME methods and the IG with the particularity of characterizing a system in terms of geometric quantities like the Ricci scalar and local scalar curvatures. It has proven to be useful in many applications \cite{geert,cafaro} and also, to study geometric phase transitions \cite{portesi}. Furthermore using ED, asymptotic expressions for information measures are obtained by means of geodesic equations \cite{cafaro}. In this geometric framework, a \emph{criterium} for characterizing global chaos can be obtained: the more negative is the curvature, the more chaotic is the dynamics. Also, local chaos can be characterized in terms of diverging initially nearby trajectories in the statistical manifold.

Besides, in dynamical systems theory, the \emph{ergodic hierarchy} (EH) characterizes the chaotic behavior in terms of a type of correlations between subsets of the phase space \cite{berko,lasota}. In the asymptotic limit of large times, the EH establishes that the dynamics is more chaotic when the correlation decays faster.
According to correlation decay, the four levels of EH are, from the weakest to the strongest: ergodic, mixing, Kolmogorov, and Bernoulli.
In particular, in mixing systems any two subsets enough separated in time can be considered as ``statistically independent" which allows to use an statistical description of the behavior of the system.
In the context of quantum chaos, the statistical independence is present in the universal statistical properties of energy levels which are given by the Gaussian ensembles \cite{stockmann}. In Gaussian ensembles theory one assumes that in a fully chaotic quantum system the interactions are neglected in such way that the Hamiltonian matrix elements can be considered statistically independent between them \cite{bgs}. Related to this, in \cite{paper1,paper2} a quantum extension of the EH was proposed, called the \emph{quantum ergodic hierarchy}, which allowed to characterize the chaotic behaviors of the Casati-Prosen model \cite{casati} and the kicked rotator \cite{stockmann}.

Inspired by the characterizations of quantum chaotic systems made in \cite{paper1,paper2} we discuss the $2\times 2$ \emph{Gaussian Orthogonal Ensembles} (GOE) within a $2D$ correlated statistical model and we propose a generalization of the ergodic, mixing and Bernoulli levels of the EH in the context of the IG and ED, called the \emph{information geometric ergodic hierarchy} (IGEH).
In order to study the correlation decay within IGEH we also define a distinguishability measure for a $2D$ correlated model that allows to give an upper bound for the information geometric correlation.

In this way, our main contribution is the information geometric ergodic hierarchy as an alternative framework for studying the chaotic dynamics in curved statistical models.
The present work is organized as follows. First, we describe the GOE in terms of a $2D$ correlated model, including a brief discussion about the dynamics characterized by the global chaos criteria. Second, we propose an information geometric extension of the ergodic hierarchy by expressing the correlations in terms of probability distributions instead of subsets of phase space. Next, we define a distinguishability measure for the $2D$ correlated model and an upper bound for the IGEH correlation is given. Finally, we draw some conclusions and future research directions are outlined.

\section{GAUSSIAN ORTHOGONAL ENSEMBLES WITHIN A CURVED STATISTICAL MODEL}

In Gaussian Orthogonal Ensembles theory one deals with the probability distribution $p(H_{11},H_{12},\ldots,H_{NN})$ for the Hamiltonian matrix elements assuming that the $H_{ij}$ are uncorrelated. Then in the framework of information geometry one could try to describe them by defining
an appropriate \emph{microspace} $\{(x_1,x_2,\ldots,x_N)\}$ that is the set of variables of the system and, a \emph{macrospace}
$\{(\theta_1,\theta_2,\ldots,\theta_M)\}$
which represents macroscopic quantities that can be measured in an experiment, like the mean value of any variable $\langle x_j\rangle$, etc.

In order to characterize the GOE within a statistical model we study a correlated ensemble of $2\times 2$ matrices.
We take the microspace as the Hamiltonian matrix elements $\{H_{11},H_{12},H_{21},H_{22}\}$ and define the macrospace as follows.
By the sake of simplicity, we choose the macrospace in such way that only $H_{11}$ and $H_{22}$ are correlated, and that the mean values of all variables are zero. Except the corresponding to $H_{11}$ which is equal to $\mu$. Also, we consider that the variance of $H_{11}$, $H_{12}$ and $H_{21}$ are the same, denoted by $\sigma$. Moreover, in order to study how independent the diagonal Hamiltonian elements are, we restrict the dynamics by considering that $r\in[-1,1]$ is the correlation coefficient between $H_{11}$ and $H_{22}$, and that the product of the covariances between $H_{11}$ and $H_{22}$ is a constant $\Sigma^2$.
Taking this into account, the resulting macrospace is $\{(\mu,\sigma)\in \mathbb{R}\times\mathbb{R}_{+}\}$ where the constraints over the variables are
\begin{eqnarray}\label{2-1}
&\int p(H_{11},H_{22},H_{12},H_{21}) \ H_{11} \ dH_{11}dH_{12}dH_{21}dH_{22}=\mu \ \ \ \  \nonumber \\
& \nonumber \\
&\int p(H_{11},H_{22},H_{12},H_{21}) \ H_{12} \ dH_{11}dH_{12}dH_{21}dH_{22}=\int p(H_{11},H_{22},H_{12},H_{21}) \ H_{21} \ dH_{11}dH_{12}dH_{21}dH_{22}= \ \ \ \  \nonumber \\
&=\int p(H_{11},H_{22},H_{12},H_{21}) \ H_{22} \ dH_{11}dH_{12}dH_{21}dH_{22}=0 \ \ \ \  \nonumber \\
& \nonumber \\
&\int p(H_{11},H_{22},H_{12},H_{21}) \ (H_{11}-\mu_{11})^2 \ dH_{11}dH_{12}dH_{21}dH_{22}=p(H_{11},H_{22},H_{12},H_{21}) \ H_{12}^2 \ dH_{11}dH_{12}dH_{21}dH_{22}= \nonumber\\
&=\int p(H_{11},H_{22},H_{12},H_{21}) \ H_{21}^2 \ dH_{11}dH_{12}dH_{21}dH_{22}=\sigma^2 \nonumber \\
& \nonumber \\
&\int p(H_{11},H_{22},H_{12},H_{21}) \ H_{22}^2 \ dH_{11}dH_{12}dH_{21}dH_{22}=\Sigma^4/\sigma^2  \ \ \ \ ,\ \ \ \ \Sigma=constant \nonumber\\
& \nonumber \\
&\int p(H_{11},H_{22},H_{12},H_{21}) \ (H_{11}-\mu)H_{22} \ dH_{11}dH_{12}dH_{21}dH_{22}=r\Sigma^2 \ \ \ \ ,\ \ \ \ -1\leq r\leq 1
\end{eqnarray}
with the normalization condition
\begin{eqnarray}\label{2-2}
\int p(H_{11},H_{22},H_{12},H_{21})dH_{11}dH_{12}dH_{21}dH_{22}=1
\end{eqnarray}
In the GOE the Hamiltonian matrix is assumed to be symmetric $H_{21}=H_{12}$.
According to MaxEnt, the probability distribution $p(H_{11},H_{22},H_{12},H_{21}|\mu,\sigma;r)$ that maximizes the Shannon--Gibbs entropy subjected to (\ref{2-1}) and (\ref{2-2}) is \cite{cafaro}
\begin{eqnarray}\label{2-3}
&p(H_{11},H_{22},H_{12},H_{21}|\mu,\sigma;r)=\nonumber\\
&\frac{1}{2\pi\Sigma^2\sqrt{1-r^2}}\exp\left(-\frac{1}{2(1-r^2)}\left[\frac{1}{\sigma^2}({H_{11}-\mu)^2}+\frac{\sigma^2}{\Sigma^4}H_{22}^2-
2r\frac{1}{\Sigma^2}(H_{11}-\mu)H_{22}\right]\right)
\frac{1}{2\pi\sigma^2}\exp\left(-\frac{1}{2\sigma^2}(H_{12}^2+H_{21}^2)\right)
\end{eqnarray}
where the correlation coefficient is considered as an external parameter and, therefore, only $\mu$ and $\sigma$ are the macrovariables of the macrospace of this model. The Fisher--Rao metric of a given statistical model is computed as
\begin{eqnarray}\label{2-4}
g_{ij}=\int dx_1\cdots dx_N p(x_1,\ldots,x_N|\theta_1,\ldots,\theta_M)\frac{\partial{\log p(\ldots|\ldots)}}{\partial \theta_i}\frac{\partial{\log p(\ldots|\ldots)}}{\partial \theta_j}  \ \ \ \ \forall \ i,j=1,\ldots M
\end{eqnarray}
where $M$ is the dimension of the macrospace and the integration is taken over all the microspace.
Since here $M=2$ with $\theta_1=\mu$ and $\theta_2=\sigma$, then, with the help of \cite{cafaro} one has
\begin{eqnarray}\label{2-5}
g_{11}=\frac{1}{\sigma(1-r^2)} \ \ \ \ , \ \ \ \ g_{22}=\frac{4}{\sigma(1-r^2)} \ \ \ \ , \ \ \ \ g_{12}=g_{21}=0
\end{eqnarray}
From this, one can obtain the Christoffel symbols $\Gamma_{ij}^k$ whose non vanishing coefficients are
\begin{eqnarray}\label{2-6}
\Gamma_{12}^1=\Gamma_{21}^1=-\frac{1}{\sigma} \ \ \ \ , \ \ \ \ \Gamma_{11}^2=\frac{1}{4\sigma} \ \ \ \ , \ \ \ \ \Gamma_{22}^2=-\frac{1}{\sigma}
\end{eqnarray}
Using (\ref{2-5}) and (\ref{2-6}) one can calculate the non vanishing components of the Ricci tensor $R_{ij}$ and the Ricci scalar curvature $R$, thus resulting
\begin{eqnarray}\label{2-7}
R=g^{11}R_{11}+g^{22}R_{22}=-\frac{1}{2}(1-r^2) \ \ \ \ \ \ \  \textrm{with} \ \ \ \ \ \ \  R_{11}=-\frac{1}{4\sigma^2} \ , \ R_{22}=-\frac{1}{\sigma^2}
\end{eqnarray}
Three remarks follow. First, the statistical manifold has a curvature which is \emph{globally} negative
for all values of the correlation coefficient $r\in[-1,1]$. Based on the global chaos \emph{criterium} above, this simply means that the dynamics in macrospace $(\mu,\sigma)$ is chaotic for all $r$.

Second, the $2\times 2$ GOE case corresponds to $r=0$ and $\Sigma=\sigma$, thus having the minimum value of the scalar curvature
\begin{eqnarray}\label{2-8}
R_{GOE}=R(r=0)=-\frac{1}{2}=R_{\min} \ \ \ \ \ \ \ \ \ \ \ \ \ \ \ \ \ \ \ \ \ \ (\textrm{GOE, more chaotic case})
\end{eqnarray}
Third, for the strongly correlated case that corresponds to $|r|\sim1$ one has
\begin{eqnarray}\label{2-9}
R(|r|\rightarrow1)=0 \ \ \ \ \ \ \ \ \ \ \ \ \ \ \ \ \ \ \ \ \ \ (\textrm{strongly correlated case})
\end{eqnarray}
which can be interpreted, by the global chaos criterium, as the case when the dynamics is the least chaotic of all.


\section{TOWARDS AN INFORMATION GEOMETRIC DEFINITION OF THE ERGODIC HIERARCHY}
Motivated by the characterization of chaotic dynamics made in \cite{paper1,paper2} by means of a quantum extension of the ergodic hierarchy, now we
study a generalization of the EH in the context of the information geometry. This allows to measure how independent the variables $H_{ij}$ are for the $2\times 2$ correlated ensemble (\ref{2-3}).

If one has a dynamical system $(X,\Sigma,\mu,\{T_t\}_{t\in J})$ where $X$ is a set, $\Sigma$ is a sigma--algebra of $X$, $\mu$ a measure defined over $\Sigma$ and $\{T_t\}_{t\in J}$ a group of preserving--measure transformations, then
the EH correlation $C(T_tA,B)$ between two subsets $A$ and $B$ of $X$ that are separated by a time $t$ is
\begin{eqnarray}\label{3-1}
C(T_tA,B)=\mu(T_tA \cap B)-\mu(A)\mu(B)
\end{eqnarray}
The ergodic, mixing and Bernoulli levels of the EH are given in terms of (\ref{3-1}) in the following way. Given two arbitrary sets $A,B\in X$, it is said that $T_t$ is
\begin{enumerate}
  \item[$\bullet$] \emph{ergodic} if
\begin{eqnarray}\label{3-1ergodic}
\lim_{T\rightarrow\infty}\frac{1}{T}\int_{0}^{T}C(T_tA,B)dt=0 \ \ \ \ \ \ ,
\end{eqnarray}
  \item[$\bullet$] \emph{mixing} if
  \begin{eqnarray}\label{3-1mixing}
\lim_{t\rightarrow\infty}C(T_tA,B)=0 \ \ \ \ \ \ ,
\end{eqnarray}
  \item[$\bullet$] \emph{Bernoulli} if
  \begin{eqnarray}\label{3-1Bernoulli}
C(T_tA,B)=0 \ \ \ \ \ \ \ \ \forall \ t\geq0
\end{eqnarray}
  \end{enumerate}
In ergodic systems the correlation vanishes ``in time average" for large times while in mixing systems $C(T_tA,B)$ vanishes for $t\rightarrow\infty$. In Bernoulli systems the correlation is zero for all times. These levels classifies the dynamics in terms of the type of decay of $C(T_tA,B)$ according to Eqns. (\ref{3-1ergodic}), (\ref{3-1mixing}) and (\ref{3-1Bernoulli}).

In order to express $C(T_tA,B)$ by means of probability distributions is more convenient to use the definition of (\ref{3-1}) in terms of distribution functions, which is given by
\begin{eqnarray}\label{3-2}
C(f\circ T_t,g)=\int_{X}(f\circ T_t)(x)g(x)dx-\int_{X}f(x)dx\int_{X}g(x)dx  \ \ \ \ \ \ \ \ \forall \ f,g \in \mathbb{L}^1(X)
\end{eqnarray}
where $f\circ T_t$ denotes the composition of $f$ and $T_t$, i.e. $f\circ T_t(x)=f(T_t(x))$ for all $x\in X$ and now the role of $A,B$ is played by the functions $f,g\in \mathbb{L}^1(X)$.

Now, in information geometry and entropic dynamics one has probability distributions $p(x,\theta)$ that depends on a set of parameters $\theta$ and the dynamics of the macrovariables $\theta$ is along the geodesics of the statistical manifold.
Moreover, in the statistical manifold the role of time variable $t$ of dynamical systems is played by a parameter $\tau$ along the geodesics. In order to introduce information geometry methods we propose the following approach by defining a correlation between functions as the macrovariables $\theta$ evolves along the geodesics.
Given $N$ functions $f(x_i)$, each one of them in terms of the variable $x_i$ for all $i=1,\ldots,N$, we propose the following \emph{IG correlation} $C(f_1,\ldots,f_N,\tau)$ between $f_1,\ldots,f_N$ at time--like parameter $\tau$ as
\begin{eqnarray}\label{3-4}
&C(f_1,\ldots,f_N,\tau)=\nonumber\\
&\int p(x_1,\ldots,x_l|\theta(\tau))\prod_{i=1}^{N}f_i(x_i)dx_1\cdots dx_l-\prod_{i=1}^{N}\int p_i(x_i|\theta(\tau))f_i(x_i)dx_i
\end{eqnarray}
where $\theta(\tau)=(\theta_1(\tau),\ldots,\theta_M(\tau))$ is the M--dimensional vector of the macrovariables at ``time" $\tau$ and,
\begin{eqnarray}\label{3-5}
p_i(x_i|\theta(\tau))=\int p(x_1,\ldots,x_N|\theta(\tau))\prod_{j\neq i}dx_{j} \ \ \ \ \ , \ \ \ \ i=1\ldots N
\end{eqnarray}
are the marginal distributions of $p(x_1,\ldots,x_N|\theta(\tau))$. From (\ref{3-4}) we can see that
$C(f_1,\ldots,f_N,\tau)$ measures how independent the variables $x_1,\ldots,x_N$ are at $\tau$, and this can be considered as a sort of information geometric generalization of the EH correlation.

Having established $C(f_1,\ldots,f_N,\tau)$ and taking into account the ergodic, mixing and Bernoulli levels given by Eqns. (\ref{3-1ergodic}), (\ref{3-1mixing}) and (\ref{3-1Bernoulli}), we define the \emph{information geometric ergodic hierarchy} (IGEH) as follows. Given a set of $N$ arbitrary functions $f_1(x_1),\ldots,f_N(x_N)$ we say that now the statistical model is
\begin{enumerate}
  \item[$\bullet$] \emph{IG ergodic} if
\begin{eqnarray}\label{3-5ergodic}
\lim_{T\rightarrow\infty}\frac{1}{T}\int_{0}^{T}C(f_1,\ldots,f_N,\tau)d\tau=0 \ \ \ \ \ \ ,
\end{eqnarray}
  \item[$\bullet$] \emph{IG mixing} if
  \begin{eqnarray}\label{3-5mixing}
\lim_{\tau\rightarrow\infty}C(f_1,\ldots,f_N,\tau)=0 \ \ \ \ \ \ ,
\end{eqnarray}
  \item[$\bullet$] \emph{IG Bernoulli} if
  \begin{eqnarray}\label{3-5Bernoulli}
C(f_1,\ldots,f_N,\tau)=0 \ \ \ \ \ \ \ \ \forall \ \tau\in \mathbb{R}
\end{eqnarray}
  \end{enumerate}
As an example, taking $r=0$ and $\Sigma=\sigma$ in (\ref{2-3}) we obtain the $2\times 2$ GOE probability distribution which is simply the product of its marginal distributions. Then from (\ref{3-4}) and (\ref{3-5Bernoulli}) it follows that the $2\times 2$ GOE is a statistical model that is IG Bernoulli.

A statistical model that is IG ergodic can be given assuming that $C(f_1,\ldots,f_N,\tau)$ is, for instance, proportional to $\sin(\alpha\tau)||f_1||_1\ldots||f_N||_1$ with $\alpha\in\mathbb{R}$. Replacing this in (\ref{3-5ergodic}) one obtains that $\lim_{T\rightarrow\infty}\frac{1}{T}\int_{0}^{T}C(f_1,\ldots,f_N,\tau)d\tau=0$. Since $\sin(\alpha\tau)||f_1||_1\ldots||f_N||_1$ oscillates then this model is not IG mixing nor IG Bernoulli.

Our approach, thus, deals with ergodic hierarchy in statistical models from an information geometry viewpoint.
Henceforth, we call \emph{ergodic statistical model} to any statistical model satisfying (\ref{3-5ergodic}), (\ref{3-5mixing}) or (\ref{3-5Bernoulli}).


\section{A DISTINGUISHABILITY MEASURE FOR THE $2D$ CORRELATED MODEL}
In order to use the levels of the IGEH to characterize the dynamics of statistical models one should have a manner of determining the decay of correlation $C(f_1,\ldots,f_N,\tau)$ according to some of the Eqns. (\ref{3-5ergodic}), (\ref{3-5mixing}) or (\ref{3-5Bernoulli}).
For the the family of the $2D$ correlated probabilities $p(H_{11},H_{22},H_{12},H_{21}|\mu,\sigma;r)$ of Eq. (\ref{2-3}) we define the following distinguishability measure $F:\{p(H_{11},H_{22},H_{12},H_{21}|\mu,\sigma;r):\mu\in(-\infty,\infty),\sigma\in(0,\infty),-1\leq r\leq1\}\longmapsto \mathbb{R}$, given by
\begin{eqnarray}\label{4-1}
&F(p)=\|p(H_{11},H_{22},H_{12},H_{21}|\mu,\sigma;r)-p(H_{11})p(H_{22})p(H_{12})p(H_{21})\|_{\infty}\nonumber\\
&=\max_{(H_{11},H_{22},H_{12},H_{21})\in \mathbb{R}^4}|p(H_{11},H_{22},H_{12},H_{21}|\mu,\sigma;r)-p(H_{11})p(H_{22})p(H_{12})p(H_{21})|
\end{eqnarray}
where $p(H_{ij})$ with $i,j=1,2$ are the marginal distributions of $p(H_{11},H_{22},H_{12},H_{21}|\mu,\sigma;r)$.
Furthermore, if $f_1(H_{11}),f_2(H_{22}),f_3(H_{12}),f_4(H_{21})\in \mathbb{L}^1(\mathbb{R})$ are arbitrary functions of $H_{11},H_{22},H_{12},H_{21}$, then we have
\begin{eqnarray}\label{4-2}
&|C(f_1,f_2,f_3,f_4,\tau)|\nonumber\\
&=\left|\int d^4 H_{ij} p(H_{11},H_{22},H_{12},H_{21}|\mu,\sigma;r)f_1(H_{11})f_2(H_{22})f_3(H_{12})f_4(H_{21})-\langle f_1(H_{11})\rangle\langle f_2(H_{22})\rangle\langle f_3(H_{12})\rangle\langle f_4(H_{21})\rangle\right|\nonumber\\
&\leq \max_{(H_{11},H_{22},H_{12},H_{21})\in \mathbb{R}^4}|p(H_{11},H_{22},H_{12},H_{21}|\mu,\sigma,r)-p(H_{11})p(H_{22})p(H_{12})p(H_{21})| \nonumber\\
&\times \left|\int d^4 H_{ij} f_1(H_{11})f_2(H_{22})f_3(H_{12})f_4(H_{21})\right|\leq F(p)||f_1||_{\infty} \ ||f_2||_{\infty} \ ||f_3||_{\infty} \ ||f_4||_{1} \
\end{eqnarray}
where $d^4 H_{ij}$ is the volume element $dH_{11}dH_{22}dH_{12}dH_{21}$ and
\begin{eqnarray}
&\langle f_1(H_{11})\rangle=\int p(H_{11})f_1(H_{11})dH_{11} \nonumber\\
&\langle f_2(H_{22})\rangle=\int p(H_{22})f_2(H_{22})dH_{22} \nonumber\\
&\langle f_3(H_{12})\rangle=\int p(H_{12})f_3(H_{12})dH_{12}\nonumber\\
&\langle f_4(H_{21})\rangle=\int p(H_{21})f_4(H_{21})dH_{21}\nonumber
\end{eqnarray}
Eq. (\ref{4-2}) expresses that $F(p)$ is un upper bound for $C(f_1,f_2,f_3,f_4,\tau)$. Therefore, it is convenient to find an analytic expression for (\ref{4-1}). After some algebra one can obtain that
\begin{eqnarray}\label{4-3}
F(p)=|r|\left(\sqrt{1-r^2}(1+|r|)\right)^{-1-\frac{1}{|r|}} \ \ \ \ \ \ \forall \ \ r\in[-1,1]
\end{eqnarray}
\begin{figure}[h]\label{fig}
  \centerline{\includegraphics[width=250pt]{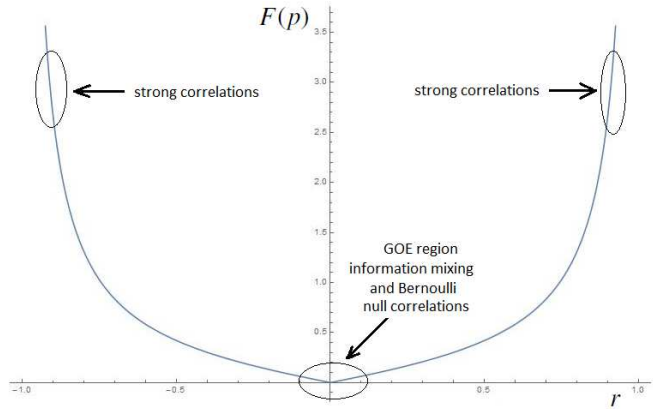}}
  \caption{$F(p)$ in function of the correlation coefficient $r$. The Gaussian Orthogonal Ensembles corresponds to the region near to $r=0$ where the statistical model belongs to the IG Bernoulli level. When $r\rightarrow\pm1$ one has that $F(p)$ diverges with the presence of strong correlations.}
\end{figure}
The behavior of $F(p)$ is shown in Fig. \ref{fig} which is independent of $\mu$ and $\sigma$. Two relevant regions, corresponding to the limiting cases $r\rightarrow0$ and $r\rightarrow\pm1$, can be well distinguished. The region $r\rightarrow0$ corresponds to the zone near the GOE which is characterized by the IG mixing and IG Bernoulli levels, with the particularity that the variables of microspace are uncorrelated. Moreover, we can see that near to $r=0$ the decay is linear in $r$. The curve $F(p)$ also shows that, if $r\rightarrow0$ when $\tau\rightarrow\infty$, then the statistical model is IG mixing.

In the region $r\rightarrow\pm1$ the measure $F(p)$ diverges corresponding to the maximally correlated case, which physically means that the system presents strong correlations between the variables of microspace. Due the correlations are strong then in this regime the statistical model can not be IG mixing nor IG Bernoulli.

Finally, it should be noted that $F(p)$ does not allow to distinguish between two probability distributions having $r$ and $-r$ respectively. The symmetry respect to the axis $r=0$ is due the mathematical form of the infinite norm $||.||_{\infty}$ in the definition (\ref{4-1}). That is, with other choices of $F(p)$ one could intend to distinguish states (probability distributions) with $r$ and $-r$.



\section{CONCLUSIONS}
We proposed a generalization of the ergodic, mixing and Bernoulli levels in the context of the information geometry called the information geometric ergodic hierarchy (IGEH), and we applied it to characterize the $2\times 2$ Gaussian Orthogonal Ensambles within a $2D$.

By defining a measure for the family of the $2D$ correlated probability distributions we obtained an upper bound for the IG correlation
which allowed us to give a necessary condition for the IG mixing level when $\lim_{\tau\rightarrow\infty}r(\tau)=0$, being $r$ the correlation coefficient and $\tau$ an arbitrary parameter along the geodesics.

The relevance of our main contribution, the information geometric ergodic hierarchy (IGEH), lies in the following remarks:
\begin{itemize}
\item The IGEH generalizes the notions of statistical independence and chaos characterization of the ergodic hierarchy to statistical models on curved manifolds in the context of the information geometry. In turn, this gives place to an ergodic hierarchy characterization of statistical models, that we called \emph{ergodic statistical models}.
\item Geometrical notions and global chaos \emph{criterium} of entropic dynamics can be related with the levels of the IGEH. The $2\times 2$ GOE case belonging to the most chaotic level, the IG Bernoulli, has associated a maximum negative value of the scalar curvature $R_{GOE}=-\frac{1}{2}$.
\item By obtaining upper bounds $F(p)$ of the IG correlation for an specific family of probability distributions, as depicted by the curve of Fig. \ref{fig}, one could study geometrical phase transitions moving along curves $F(p)$ as an external parameter $r$ is varied.
\end{itemize}


\section{ACKNOWLEDGMENTS}
This work was partially supported by CONICET (National Research Council), the Universidad Nacional de La Plata and the Instituto de F\'{i}sica La Plata, La Plata, Argentina.


\nocite{*}
\bibliographystyle{aipnum-cp}%

\end{document}